# Fine structure in the Sigma Orionis cluster revealed by *Gaia* DR3

M. Žerjal[1,2]★, E. L. Martín[1,2], and A. Pérez-Garrido[3]

[1] Instituto de Astrofísica de Canarias, E-38205 La Laguna, Tenerife, Spain
[2] Universidad de La Laguna, Dpto. Astrofísica, E-38206 La Laguna, Tenerife, Spain
[3] c, 30202 Cartagena, Murcia, Spain

April 25, 2024

**ABSTRACT**

*Context.* Sigma Orionis is an open cluster in the nearest giant star formation site – Orion. Its youth (3-5 Myr), low reddening, and relative vicinity make it an important benchmark cluster to study stellar and substellar formation and evolution.
*Aims.* Young star-forming sites are complex and hierarchical. Precision astrometry from *Gaia* DR3 enables the exploration of their fine structure.
*Methods.* We used the modified convergent point technique to kinematically re-evaluate the members in the Sigma Orionis cluster and its vicinity.
*Results.* We present clear evidence for three kinematically distinct groups in the Sigma Orionis region. The second group, the RV Orionis association, is adjacent to the Sigma Orionis cluster and is composed only of low-mass stars. The third group, the Flame association, whose age is comparable to that of Sigma Orionis, overlaps with the younger NGC 2024 in the Flame Nebula. In total, we have discovered 105 members of this complex not previously found in the literature (82 in Sigma Orionis, 19 in the Flame association, and 4 in the RV Orionis association).

## 1. Introduction

Sigma Orionis ($\sigma$ Ori) is a benchmark open cluster in the nearest giant star formation complex – Orion. There are multiple dark clouds and stellar populations in Orion, indicating a complicated history of star formation over the last 15 Myr (Kubiak et al. 2017; Kounkel et al. 2018; Zari et al. 2019). The $\sigma$ Ori cluster itself is located in the Orion C region, which is relatively clear of molecular cloud material and where there is no ongoing star formation. Close to it, however, there are heavily obscured regions, such as the Horsehead Nebula (Barnard 33). North of $\sigma$ Ori are the two Orion OB1 subgroups of OB stars identified by Blaauw (1964). It has been suggested that the low-mass population of OB1b overlaps with that of the northern part of the $\sigma$ Ori cluster, but the two populations have different distances and radial velocities (Jeffries et al. 2006; Maxted et al. 2008).

The existence of a clustering of B-type stars and late-type young stars around the bright O9.5 star $\sigma$ Ori was recognised by Garrison (1967) and Walter et al. (1997), respectively. Low-mass stars were identified using the Röntgensatellit (ROSAT) X-ray observations and follow-up optical imaging and spectroscopy. Very low-mass candidate members straddling the stellar–substellar border have been found using wide-area optical and infrared imaging and spectroscopy (Béjar et al. 1999; Zapatero Osorio et al. 2002), and deep imaging and reconnaissance spectroscopic observations have even reached the planetary-mass domain (Zapatero Osorio et al. 2000; Barrado y Navascués et al. 2001; Martín et al. 2001; Peña Ramírez et al. 2012). A pre-*Gaia* review of the properties of the $\sigma$ Ori cluster can be found in Walter et al. (2008).

Detailed membership studies using *Gaia* are essential because the $\sigma$ Ori cluster is located in a complex region. *Gaia* Data Release (DR) 2 astrometric data were used in the large spectroscopic membership study by Caballero et al. (2019). They report a significant number of unconfirmed cluster members among the list of members provided by previous studies. This shows the importance of including astrometric and kinematic information when determining cluster memberships.

The main aim of this paper is to re-assess the membership of the cluster and its subcomponents along the entire mass range available in the *Gaia* DR3 catalogue. Typical young cluster membership indicators (colour-magnitude diagram, H$\alpha$ emission, and Li I and Na I absorption) are examined but are not used to determine membership. The structure of this paper is as follows. We describe the input catalogue and the mass estimation in Section 2. Section 3 includes a discussion of the division of the complex star-forming region into subgroups and their membership determination. We comment on the revised membership lists and the fine substructure of the $\sigma$ Ori region in Section 4 and supplement our membership lists with the spectroscopic indicators. We summarize our findings and conclude in Section 5.

## 2. Data

Our membership selection is based on stellar kinematics. We used the stellar coordinates, parallaxes, proper motions, and radial velocities in the *Gaia* DR3 catalogue (Gaia Collaboration et al. 2016, 2022). We computed distances as `1/parallax`. As estimated by Žerjal et al. (2023), the difference between the distances based on the Bayesian approach (Bailer-Jones et al. 2021) and the inverted parallax is small at the distance of Orion, on average less than 1 pc within 400 pc and 4 pc beyond this limit.

★ E-mail: mzerjal@iac.es

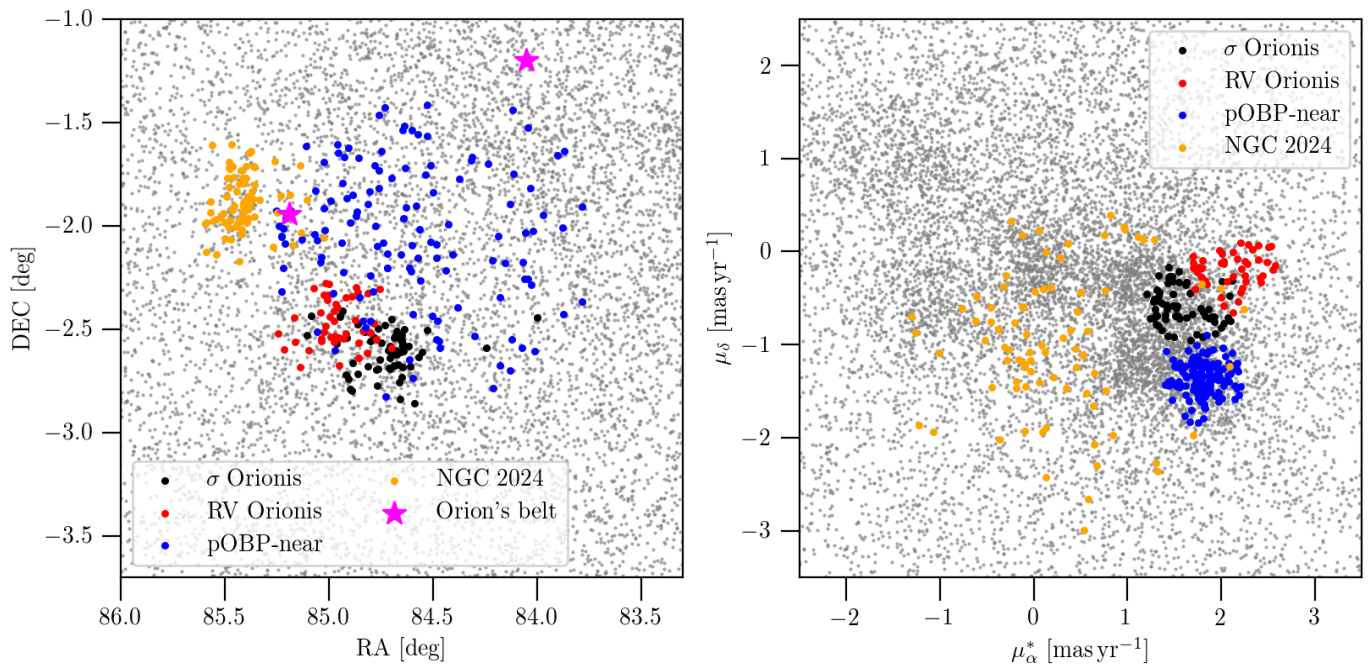

**Fig. 1.** Input catalogue (grey dots) and the positions of the preliminary members of the four different overdensities in the $\sigma$ Ori region. The clusters $\sigma$ Ori and RV Ori show overdensities in the sky as well as in the proper motion space. On the other hand, the pOBP-near component is dense in proper motion space but sparse in the sky, while the opposite is true in the Flame association.

Our selection criteria for the initial source list were based on coordinates ($80 < \alpha < 90$ deg, $-4 < \delta < 0$ deg) and parallax ($2 < \pi < 5$ mas). This is a relatively wide cut to ensure that no potential members are left out. We did not perform any quality cuts in the data in order to reach the faintest cluster members. In total, 14% of stars have radial velocities in *Gaia* (among them almost 8% of K and 2% of M dwarfs).

The brightest star in the cluster, $\sigma$ Orionis AB (one single entry *Gaia* DR3 3216486443742786048 that is actually an unresolved triple system), is saturated in *Gaia* (G=3.4 mag) and is therefore lacking parallax and proper motion measurements. We took proper motions from the new reduction of Hipparcos data for this star despite the large measurement uncertainties ($\mu_\alpha^* = 22.63 \pm 10.83$ mas yr$^{-1}$, $\mu_\delta = 13.45 \pm 5.09$ mas yr$^{-1}$; van Leeuwen 2007). Its distance, however, is most precisely determined by interferometry ($387.5 \pm 1.3$ pc; Schaefer et al. 2016).

## 3. Membership determination

Our membership determination algorithm is based on the Perryman technique (Perryman et al. 1998). The procedure calls for the input data to be composed of two tables, one with the list of candidate members covering the entire anticipated parameter space of the cluster, and another with the list of the preliminary members that serves as a starting point to estimate the systemic velocity of the cluster. We describe all the steps in detail below.

### 3.1. Preliminary members

We manually selected the preliminary members in a cautious manner, aiming to minimize the contamination rate while ensuring an adequate number of members for accurately computing the initial barycentre and systemic velocity of the cluster. The parameter space in the $\sigma$ Ori star-forming region is intertwined with numerous distinct young populations, as depicted in Figure 1 and explored in detail by Chen et al. (2020). The $\sigma$ Ori cluster itself and its immediate vicinity consist of five kinematically distinct groups. Upon initial examination of the input catalogue, it became evident that there are two adjacent but clearly separate populations within the $\sigma$ Ori cluster in the proper motion space. Both newly identified groups consist exclusively of low-mass stars that share a similar age with the $\sigma$ Ori cluster. We named the first the RV Orionis association (RV Ori) after its most luminous member (spectral type K5.0, Hernández et al. 2014). It is located between the core of the $\sigma$ Ori cluster and the Horsehead Nebula. The second group – the Flame association – has an inconspicuous presence due to a low number of stars and its overlap with NGC 2024. Additionally, we characterised NGC 2024, which is embedded in the Flame Nebula (Levine et al. 2006; Getman et al. 2014) and younger than the $\sigma$ Ori cluster (< 1 Myr, Kuhn et al. 2015), in order to effectively disentangle it from the main group.

The fifth population is a sparse and slightly older group in front of the $\sigma$ Ori cluster. It is seen as a populous overdensity in the proper motion space but sparse in the sky. It is related to a group historically named OB1b, and recently characterised as OBP-near by Chen et al. (2020), where OBP stands for the Orion Belt population. Due to its broader extent compared to our input catalogue, certain members of this cluster are not included in our membership list. Consequently, in this study, we designated its members as partial OBP-near (pOBP-near), clarifying that our interest lies in its limited characterisation for the purpose of distinguishing it from the main cluster, rather than conducting a comprehensive analysis because it belongs to an older star-forming event.

The criteria for the selection of the preliminary members of $\sigma$ and RV Ori, NGC 2024, and pOBP-near are individually customised. Members of the Flame association were discovered and determined later in the procedure due to its inconspicuous nature, as described in Sect. 4.3.

The preparation of the table with the preliminary members was composed of two steps due to the discrepancies between the pre-*Gaia* cluster distances in the literature and those estimated from the *Gaia* parallaxes, and due to the missing information about the newly discovered RV Ori association. The first part consisted of the determination of the centres and radii of the clusters in the positional and velocity spaces and their mean radial velocities, as listed in Table 1. If available, we took this information from the literature; otherwise, we used our own values. In the case of the $\sigma$ Ori cluster, we used the membership list from Caballero et al. (2019) due to the overlap of this cluster with other overdensities in the proper motion space. We only took objects with $\mu_{\alpha^*}$ between 0 and 3 mas yr$^{-1}$, $\mu_\delta$ between -2 and 1 mas yr$^{-1}$, and parallaxes between 2 and 5 mas to eliminate the obvious outliers in their list. For RV Ori, NGC 2024, and pOBP-near, the selection was performed manually in TOPCAT (Taylor 2005) by tracing the overdensities in the proper motion and distances, and their location in the sky. These lists allowed the determination of the cluster centres and radii, as listed in Table 1. Radii in the physical and proper motion space were determined by a visual inspection and were chosen to incorporate the majority of the stars in the overdensity. Since these were preliminary lists, we aimed to achieve low contamination rather than high completeness levels.

In the second step, we applied the filters from Table 1 to produce the preliminary lists of members for each group. We centred the fields on the commonly reported ($\alpha$, $\delta$) coordinates from the literature (e.g. Caballero et al. (2019); Chen et al. (2020) for the $\sigma$ Ori and pOBP-near; Getman et al. 2014 for NGC 2024) or determined in the first step of the procedure (RV Ori). The spatial selection of stars was done within the angle $\beta$ in the sky. This angle varies from cluster to cluster and corresponds to a radius of 5 pc at its distance; this is a typical tidal radius for a young cluster within 500 pc (Žerjal et al. 2023). The distance cut-off was applied in a less restrictive manner to accommodate for uncertainties in parallax measurements. Likewise, we centred each cluster on ($\mu_{\alpha^*}$, $\mu_\delta$) in the proper motion space, and then identified objects falling within the radius $r_\mu$. This chosen radius encompassed the central overdensity and remained consistent across all clusters, except for NGC 2024, due to its sparse nature. Subsequently, we implemented the radial velocity boundaries to eliminate evident outliers. These boundaries were quite broad and primarily served to exclude obvious outliers with radial velocities that deviate significantly from typical values, for example by a factor of 3 or more with respect to the mean value for the cluster. The combination of all these cuts successfully isolated N preliminary members, the $N_{RV}$ stars of which have radial velocity measurements. This selection was very conservative and only served as a starting point for the membership selection algorithm. Figure 1 illustrates the positions of the preliminary members in the sky and the proper motion space.

### 3.2. Mass determination

The systemic velocities of the clusters were determined as the mean velocities of their members, weighted by their mass. We estimated stellar masses in the input catalogue using the colour-mass relation from the PARSEC (PAdova and TRieste Stellar Evolution Code) models (Bressan et al. 2012; Chen et al. 2014, 2015; Tang et al. 2014; Marigo et al. 2017; Pastorelli et al. 2019, 2020) for $\sim 3$ Myr (this is the cluster age estimated by Zapatero Osorio et al. 2002) and solar metallicity. We fitted a ninth-order polynomial to the $(G - G_{RP})$ – mass relation, as shown in Figure 2. The model is available up to $G - G_{RP} = 1.45$ that corresponds to $\sim 0.1\,M_\odot$ (mid-M dwarfs). For the redder stars, we fixed the mass to $0.05\,M_\odot$. While this assumption is not ideal, the limited number of low-mass stars, combined with the incompleteness in *Gaia* at this distance, results in a minimal impact on the computation of the systemic velocity.

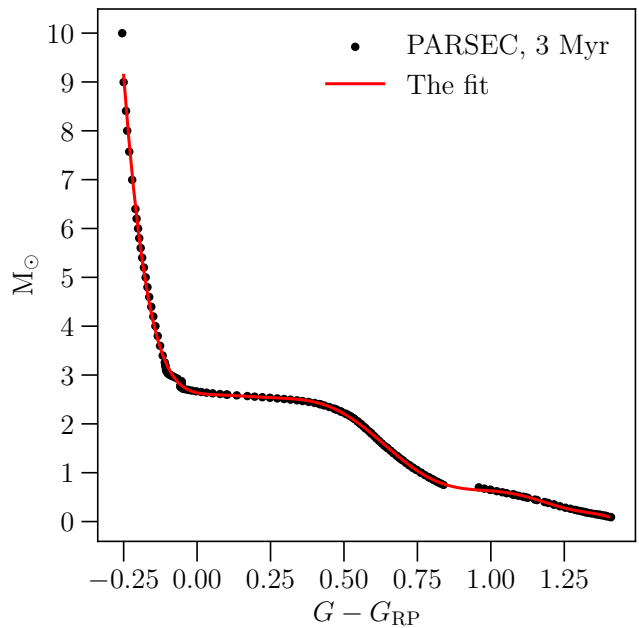

**Fig. 2.** Colour-mass relation used to estimate stellar masses. We fitted a ninth-order polynomial that is valid up to $G - G_{RP} = 1.45$, which corresponds to $\sim 0.1\,M_\odot$ (mid-M dwarfs).

The most massive member of the cluster, $\sigma$ Ori AB, is a hierarchical triple system composed of a wide pair A+B, where star A is a close spectroscopic binary composed of Aa and Ab (Simón-Díaz et al. 2011) with dynamical masses of $M_{Aa} = 16.99 \pm 0.20\,M_\odot$, $M_{Ab} = 12.81 \pm 0.18\,M_\odot$, and $M_B = 11.5 \pm 1.2\,M_\odot$ (Schaefer et al. 2016). We manually entered a total mass of $41.3\,M_\odot$ for this star.

We acknowledge the fact that our mass estimates are only approximate and that we did not account for binaries. Assuming a symmetric distribution of binaries in clusters, any potential biases in determination of the barycentre or the systemic velocity of the cluster should have cancelled out due to the large number of member stars (Perryman et al. 1998; Reino et al. 2018).

### 3.3. Membership list

To determine cluster members, we used a convergent point method developed by Perryman et al. (1998) that is based on the comparison of the kinematic properties of candidate stars with the systemic velocity of the cluster itself. We briefly summarize the method and refer the reader to the original paper (Perryman

| Cluster | $\alpha$ | $\delta$ | $\beta$ | Distance | $\mu_{\alpha*}$ | $\mu_\delta$ | $r_\mu$ | RV | N | $N_{RV}$ |
|---|---|---|---|---|---|---|---|---|---|---|
| | deg | deg | deg | pc | mas yr$^{-1}$ | mas yr$^{-1}$ | mas yr$^{-1}$ | km s$^{-1}$ | | |
| $\sigma$ Ori | 84.675 | -2.6 | 0.71 | 380 – 440 | 1.70 | -0.52 | 0.5 | [0, 60] | 64 | 18 |
| RV Ori | 84.976 | -2.41 | 0.71 | 380 – 420 | 2.12 | -0.16 | 0.5 | [0, 50] | 44 | 10 |
| pOBP-near | 84.500 | -2.09 | 0.79 | 300 – 380 | 1.78 | -1.38 | 0.5 | [-50, 50] | 127 | 39 |
| NGC 2024 | 85.444 | -1.9 | 0.73 | 360 – 430 | 0.5 | -1.5 | 2 | / | 83 | 14 |

**Notes.** Angle $\beta$ corresponds to a radius of 5 pc at the cluster's distance while $r_\mu$ is the radius in the proper motion space. N and $N_{RV}$ are the numbers of all preliminary members and those with radial velocity measurements, respectively.

**Table 1.** Selection criteria of the preliminary cluster members.

et al. 1998), but also Lodieu et al. (2019b), Lodieu et al. (2019a), and Žerjal et al. (2023) for more details.

In the first step, we determined the transversal and radial velocities of the clusters from the preliminary lists of members. The values were computed as a mean, weighted by the stellar mass. Next, we prepared the membership list by computing the transversal and radial velocities a member star would have at the location of a candidate star in question. We computed the difference between the expected and the actual velocity (vector $\mathbf{z_i}$ for the $i$-th star) and determined the $c$ value as $c = \mathbf{z}^T \Sigma^{-1} \mathbf{z}$. Here, $\Sigma$ is a confidence region that helps us estimate the scaled distance between the expected and the observed velocity vector, $\mathbf{z_i}$. It is a sum of two covariance matrices and incorporates measurement uncertainties and correlations between the observables for the measured and expected values. We can understand $c$ as a proxy for membership probability, where low $c$ values represent high membership probability. The value $c^2$ follows a $\chi^2$ distribution for the selected number of degrees of freedom. For three degrees of freedom (proper motions and radial velocity are known), a $3\sigma$ confidence interval translates to c = 14.16. If the radial velocity is not known (two degrees of freedom), $c$ = 11.83. Stars with $c$ values below these limits are considered members.

The convergence point method identifies stars whose motion in space corresponds to the velocity of the cluster. However, when dealing with close cluster pairs or complex star-forming regions where components have similar velocities, the method may struggle to accurately distinguish between the members of these nearby clusters. The determination of each cluster's membership was conducted independently using the full Gaia input catalogue, which could have lead to stars being assigned to more than one cluster. To ensure that each star is only assigned membership to one cluster, particular consideration was given to stars that appeared to have multiple memberships. In such cases, we assigned their membership to the cluster associated with the lowest $c$ value.

The membership list of pOBP-near was prepared in a slightly different way: since the algorithm is not robust enough to deal with such sparse populations, its results encompassed a wider range of proper motions and overdensities in the sky. Effectively, the first results for pOBP-near contained a large fraction of the $\sigma$ Ori members and the rest of the young groups. We solved this problem by removing the members of other clusters in this work from the pOBP-near catalogue. Similarly, we eliminated the rest of the clusters from the input catalogue for cluster NGC 2024, and implemented an exclusion criterion for stars with $\alpha < 83.5$ deg to prevent the inclusion of another nearby prominent overdensity.

To minimise the contamination rate by distant stars with larger astrometric uncertainties that appear members by chance, we introduced a radial cut from the clusters' centres based on their tidal radii. We determined the tidal radius of each cluster using Eq. 3 from Röser et al. (2011). Their relation $M_{\text{cluster}} = \frac{4A(A-B)}{G} r^3$ estimates the mass of the cluster $M_{\text{cluster}}$ within the tidal radius $r$. Oort's constants $A = 14.5$ km s$^{-1}$ kpc$^{-1}$ and $B = -13.0$ km s$^{-1}$ kpc$^{-1}$ were determined from 581 clusters within 2.5 kpc by Piskunov et al. (2006); $G$ is gravitational constant. We compared this relation with the cumulative radial mass distribution of the cluster. The intersection with the model gave us the tidal radius. Finally, we limited the volume of each cluster to 5 tidal radii. We present the list of members for the analysed populations in Appendix A and evaluate it in Section 4.

## 4. Discussion

Our membership analysis has confirmed the presence of three distinct populations in the $\sigma$ Ori group, consisting of 251 members in total. The precision of the proper motion measurements has unveiled that $\sigma$ Ori is not as homogeneous as previously believed. Instead, it consists of two distinct parts, with the second part being RV Ori. RV Ori is spatially separated from the rest of the cluster and exhibits similar but distinguishable proper motions, suggesting that it represents its own distinct population. The third component within 10 pc is the Flame association, which overlaps with the younger Flame Nebula cluster NGC 2024 but has the same age as the $\sigma$ Ori cluster. Below, we provide a description of each component.

### 4.1. $\sigma$ Orionis

Cluster $\sigma$ Ori has traditionally been seen as a homogeneous young population placed behind a sparse and slightly older group that is part of OB 1b (OBP-near). Thanks to precision astrometry from *Gaia*, we were able to split this cluster into the main part and the less populous RV Ori association that is described in the next section.

We identified 217 objects in the $\sigma$ Ori cluster in the volume of its 5 tidal radii, 82 of them are new and not listed in any known catalogues. Most of the members are concentrated in the centre that is surrounded by a sparse halo as shown in Figure 3. On average, the core and the halo show slightly different proper motions (1.488, −0.617) mas yr$^{-1}$ and (1.374, −0.915) mas yr$^{-1}$, respectively, but share the rest of the properties.

The colour-magnitude diagram in Figure 4 revealed a tight young sequence that is confined within the 3-5 Myr isochrones[1].

---
[1] We assumed negligible extinction for the $\sigma$ and RV Ori region (e.g. Oliveira et al. 2004). Stars embedded in the Flame Nebula region might be affected by a strong extinction ($A_V \sim 20$ at the core and $A_V \sim 6$ in the periphery; Getman et al. 2014). There are only a few members in the immediate vicinity of the Horsehead Nebula.

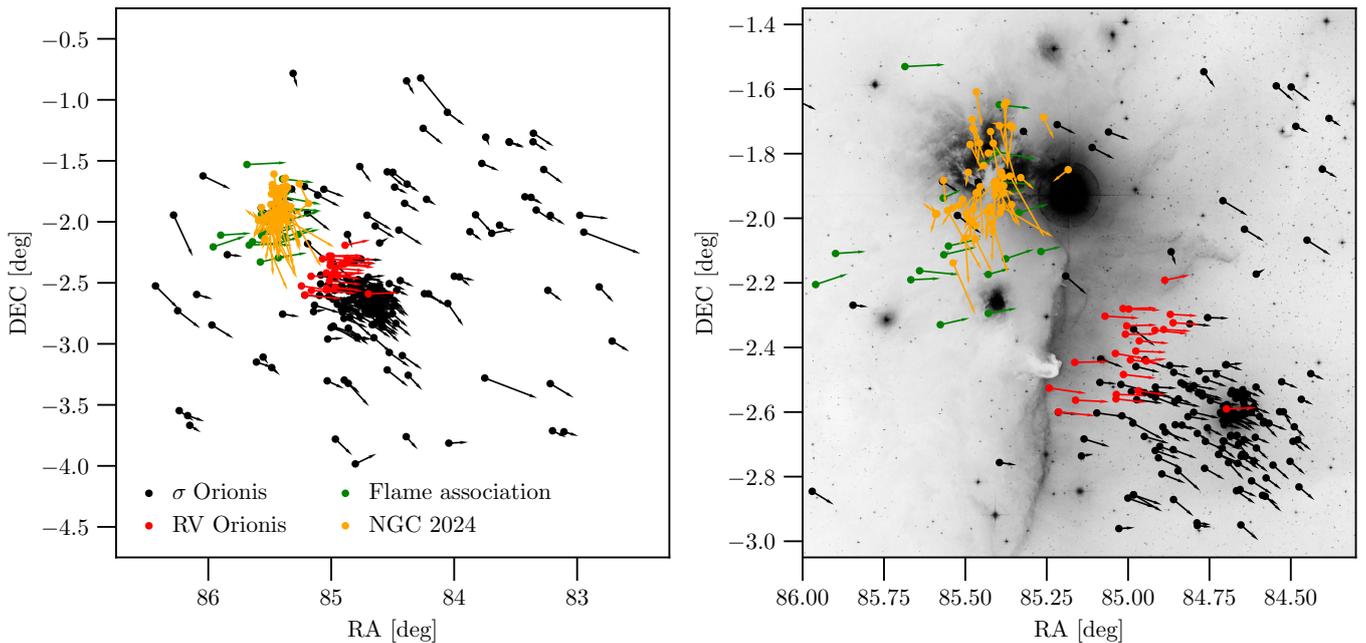

**Fig. 3.** The $\sigma$ Orionis star-forming region in the sky. It comprises three prominent subcomponents (the $\sigma$ Ori cluster, the RV Ori association, and the Flame association) as well as a halo. The Flame Nebula Cluster (NGC 2024) is located in the same region but appears younger. The right panel shows the central area of the region on top of the Digitized Sky Survey image from Aladin (Boch & Fernique 2014). Each of the populations has its distinct transversal velocities (shown in arbitrary units).

The population is composed of nine OBA stars, one F and two solar-type stars; the rest are low-mass objects. There are five AFG stars below the 5 Myr isochrone, and four of them are located in the outskirts of the cluster. Three of these are found in the direction of the Flame Nebula; their potential high extinction might explain their position in the diagram. We designated all five stars as tentative members.

This cluster was believed to be located at the distance of 388 pc (e.g. Caballero et al. 2019) and centred around the star $\sigma$ Ori. In Figure 5 we show that the majority of members are found beyond that distance, roughly between 390 and 415 pc, with a median value of 402 pc and with a spread of 9 pc. On the other hand, the distance to the star $\sigma$ Ori AB measured by interferometry is $387.5 \pm 1.3$ pc (Schaefer et al. 2016). This is contrary to our expectations; we expected to find such a massive star in the centre of the cluster. This interferometric parallax was computed with measurements from the Center for High Angular Resolution Astronomy array (CHARA), Navy Precision Optical Interferometer (NPOI), and Very Large Telescope Interferometer (VLTI). Based on the average systematic offset between *Gaia* DR2 and very long baseline interferometry astrometry of $-75 \pm 29 \,\mu$as as reported by Xu et al. (2019), future work should explore the potential systematic offset for the data used in this work before making any conclusions about the position of $\sigma$ Ori AB with respect to the cluster centre.

The proper motions of $\sigma$ Ori are small due to the dominant motion occurring along the radial direction. Consequently, prior to Gaia, cluster membership was primarily surveyed using radial velocities. Jeffries et al. (2006) determined the cluster radial velocity of $31.0 \pm 0.1$ km s$^{-1}$, with an external error of $\pm 0.5$ km s$^{-1}$. While we used only *Gaia*'s radial velocities in our kinematic membership determination, we additionally list values from other sources as a supplementary test for membership reliability and to search for potential outliers (in total, we added radial velocities for 14 stars from Sacco et al. 2008). We list these values in Table A.1. Our median values of RV=30.7 km s$^{-1}$ and 31.0 km s$^{-1}$ for $\sigma$ and RV Ori, respectively, confirm the finding of Jeffries et al. (2006). All outliers in radial velocity are positioned at around $\pm 35$ km s$^{-1}$ from the mean cluster velocity. Their $c$ values are larger than 5, which makes them slightly less reliable members. Five of them have high `ruwe` parameter, which makes them potential binaries.

The star $\sigma$ Ori is the hottest star in the cluster (spectral type O9.5) and is responsible for the illumination of the nearby Horsehead Nebula Caballero (2007). As mentioned earlier, it is in fact a hierarchical triple system composed of a wide pair A+B and a close spectroscopic binary Aa and Ab (Simón-Díaz et al. 2011). Its $c$ value is 11.4 and its distance from the cluster centre is 12.7 pc.

Other well-known bright members of this cluster in the literature are $\sigma$ Ori C, D, and E. However, our analysis only found RV Ori as a member and renounced C and E. The parallax of $\sigma$ Ori E (Gaia DR3 3216486478101981056) of $2.308 \pm 0.065$ mas places it to $433^{+12.6}_{-11.9}$ pc which is almost 20 pc behind the cluster. In the case of $\sigma$ Ori C (*Gaia* DR3 3216486439450208000), the star is located at 405 pc, but its proper motion of $\mu^*_\alpha = 0.358 \pm 0.028$ mas yr$^{-1}$ and $\mu^*_\delta = -1.064 \pm 0.027$ mas yr$^{-1}$ does not agree with the cluster parameters.

The list of members includes a white dwarf, *Gaia* DR3 3216956892983762944). It is listed in the cata-logue of white dwarfs in Gaia (Gentile Fusillo et al. 2021). This star is a reliable member of the cluster ($c = 7.68$), but it is located in the outskirts in the sky, and it appears at the edge of the cluster overdensity in the proper motion space (Figure 6). Since σ Ori is a very young association, we assigned this white dwarf as a tentative member whose membership status should be examined in more detail.

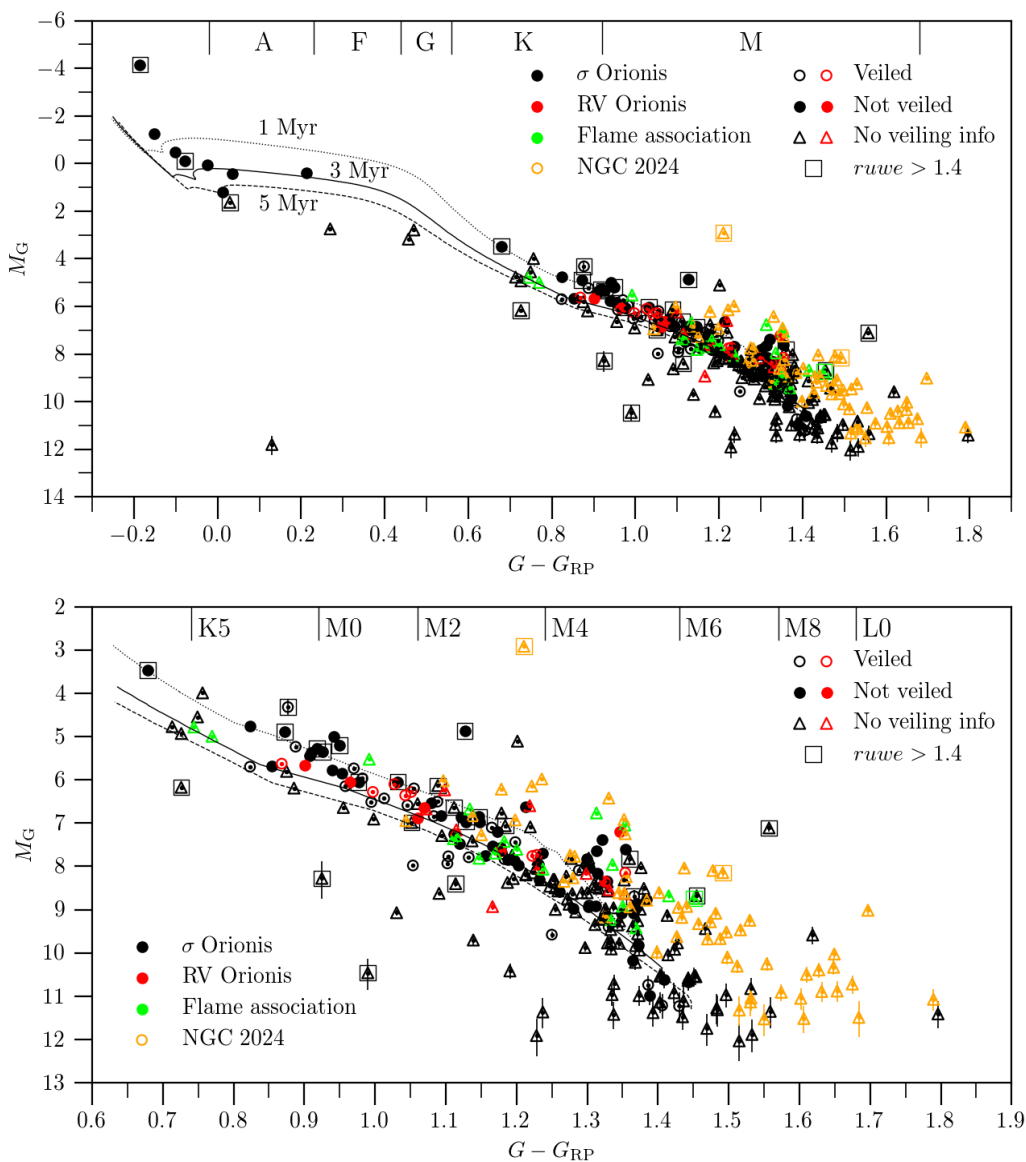

**Fig. 4.** Colour-magnitude diagram for σ Orionis, RV Orionis, and the Flame association. Yellow dots represent NGC 2024, which is affected by high extinction. The upper plot shows all members, while the bottom plot focuses only on low-mass stars. As a guideline, we added 1, 3, and 5 Myr PARSEC isochrones for [M/H]=0.

### 4.2. RV Orionis association

We identified the association of 24 stars, located adjacent to σ Ori yet distinct in the parameter space, as the RV Orionis association (RV Ori), named after its brightest member. We identified four members that were not previously recognised as part of the σ Ori cluster. This association is composed exclusively of low-mass stars, and its most luminous star, RV Ori (*Gaia* DR3 3216500531234897920), is a late-K type and exhibits an apparent *Gaia* G magnitude of 13.7 mag.

RV Ori is partially overlapping with the σ Ori cluster in the sky (Figure 3) but is clearly separated in the proper motion space as demonstrated in Figure 6. This is also reflected in the transver-sal velocities in Figure 3 that differ from those of the σ Ori members. It seems that these stars are coming from the projected direction of the nearby Horsehead Nebula that is located at the same distance as Sigma Orionis AB (e.g. Hwang et al. 2023). RV Ori is positioned at the same distance from the Sun as σ Ori (∼ 390 − 415 pc with the median value of 402 pc and a spread of 5 pc; Figure 5). While there is clear evidence that RV Ori is distinct from σ Ori, the two likely formed at the same time because both populations overlap in the colour-magnitude diagram (Figure 4).

Figure 7 reveals the consistency of the radial velocities in this association. The weighted mean for this group is $31.0 \, \text{km s}^{-1}$. Another striking observation in this plot is the fact that the majority of the members have very low c values, making them reliable members. The lack of stars with higher c values in this group comes from the fact that our membership determination method includes disentanglement of stars with double membership (see Section 3.3). It turned out that stars with higher c-values in this association more likely belong to the main σ Ori cluster.

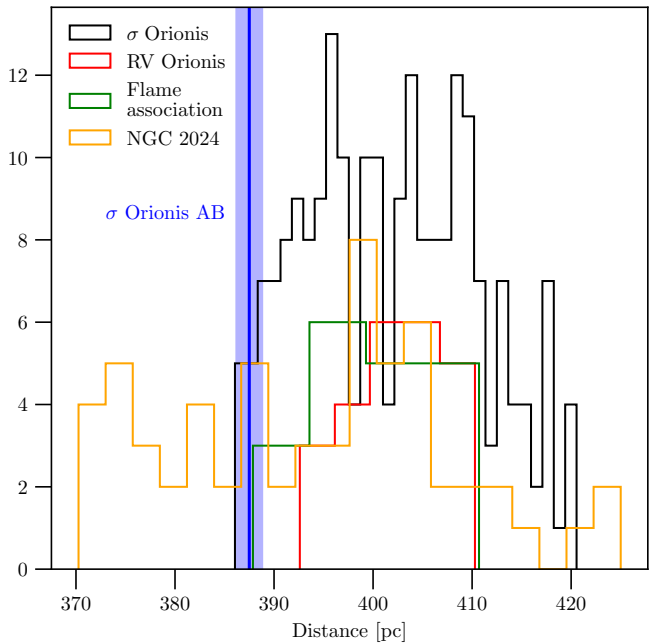

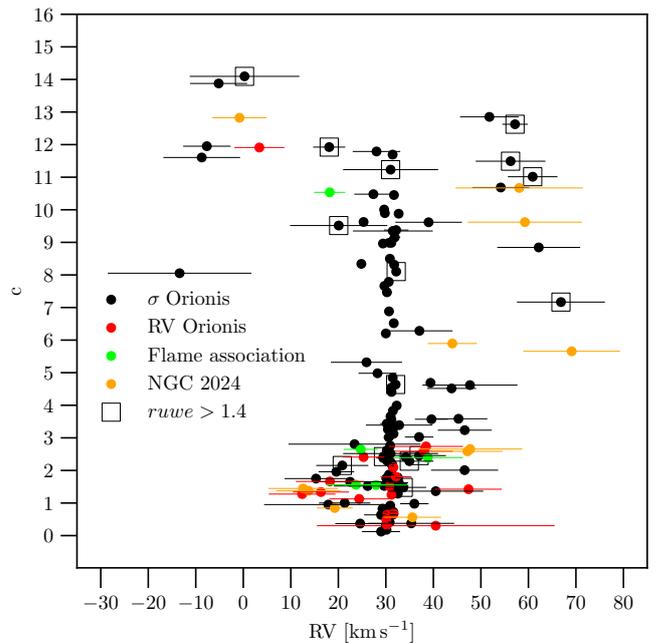

**Fig. 5.** Distance distribution for all components of the $\sigma$ Ori star-forming region. The triple star $\sigma$ Orionis AB is located in front of the cluster according to the interferometric distance from Schaefer et al. (2016).

**Fig. 7.** Radial velocity distribution versus the membership criterion, $c$. Outliers in radial velocity have relatively high $c$ values (but are still considered members) and are binary star candidates.

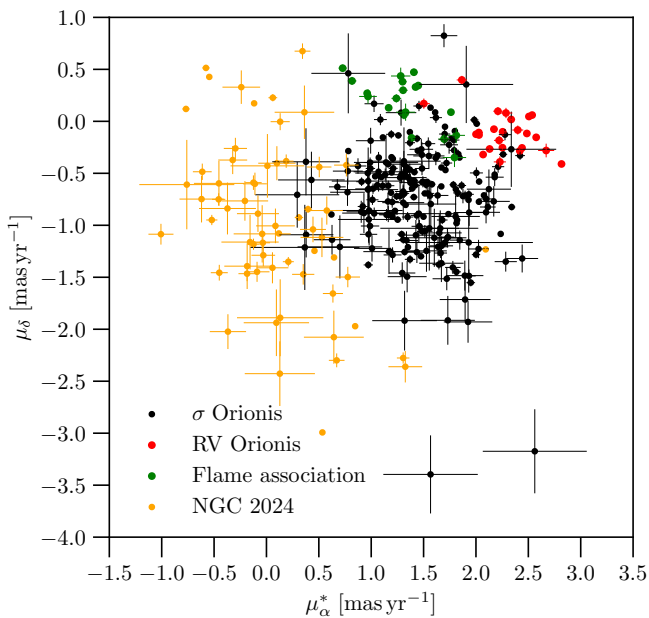

**Fig. 6.** Proper motions revealed adjacent but distinct populations in the $\sigma$ Ori region.

### 4.3. NGC 2024 and the Flame association

The Flame Nebula Cluster (NGC 2024) is seen as a strong overdensity in the sky near the star Alnitak ($\zeta$ Orionis), although they are not related. It appears very sparse in the proper motion space. Our kinematic results for this cluster and RV Ori association both revealed stars that are members but have distinct proper motions adjacent to the $\sigma$ Ori stars, and are all located in the direction of the Flame Nebula (Figure 3). Additionally, their age seems similar to the age of the $\sigma$ Ori cluster (Figure 4), while NGC 2024 is younger. In fact, Getman et al. (2014) report the age gradient in NGC 2024 with the values increasing from 0.2 Myr in the core to 1.5 Myr at the distance of ~1 pc from the centre. These stars are embedded in the Flame Nebula and affected by its strong extinction ($A_V \sim 20$ at the core and $A_V \sim 6$ in the periphery; Getman et al. 2014). We thus named the new group the Flame association. It is composed of 19 stars; all of them have low-mass and are new members of the $\sigma$ Ori star-formation site. Here we report the Flame association not only surrounding[2] the younger core of NGC 2024 with most of the members on the southern side, but also exhibiting proper motions more akin to those of $\sigma$ Ori than NGC 2024. This observation underscores the need for further investigation of the complex star-formation scenario in this region.

Our NGC 2024 consists of 62 stellar members, including the outskirts, while the catalogue from Getman et al. (2014) lists 121 objects in a smaller volume around the core. We found a crossmatch for 90 stars from their membership list in the *Gaia* catalogue. Our input catalogue contains 73 stars from this crossmatched list. Finally, we found 29 stars in common between the work of Getman et al. (2014) and our catalogue. We note that this is a highly extinct region due to the Flame Nebula, and dedicated infrared surveys like the MYStIX project described in Getman et al. (2014) are more suitable to achieve higher detection rate.

### 4.4. pOBP-near

Our pOBP-near occupies a large volume in the sky, while it is seen as a strong overdensity in the proper motion space. The majority of its members in our volume are located in front of the σ Ori cluster. While pOBP-near is clearly older, the distance distribution of both populations shows that they are located in the proximity of each other. One group seemingly extends into another. This situation could potentially lead to a partial membership confusion, where a small portion of σ Ori members might inaccurately be attributed to pOBP-near, and vice versa. However, this is likely not common. We discuss contamination in Section 4.6. We note again that our list of OBP-near members is not complete (thus a proponent 'p'). We partially characterised it in this work in order to disentangle it from the σ Ori complex and reduce the contamination rate.

---

[2] In the *Gaia* catalogue, there are no stellar detections in the cluster core due to a very high extinction. The absence of stars in the most obstructed regions is clearly evident in Figure 3.

### 4.5. Spectroscopic indicators of a young age

We prepared a compilation of the existing spectroscopic data to support their membership status. The table of cluster members with spectroscopic measurements is presented in Appendix C. There are 96 stars with measured equivalent width of Li I 6708Å line, 106 stars with Hα, and 80 stars with Na I doublet (8183 and 8195 Å).

Very young stars might be affected by non-photospheric veiling coming from hot boundary layers in active accretion discs that can diminish the equivalent widths of absorption lines (Basri et al. 1991). To identify stars that might be affected by veiling, we applied the chromospheric criterion from Barrado y Navascués & Martín (2003) that is based on the strength of Hα emission. The details are described in Appendix B, and the veiling flags are given in Table C.1. We show equivalent widths of Hα for members in Figure 8, distinguishing the veiled and non-veiled stars for RV and σ Orionis (no Hα measurements are available for other populations). Interestingly, among the stars with equivalent width Hα measurements, RV Ori exhibits a higher proportion of veiled stars (53%) than σ Ori (32%). On the other hand, their colour-magnitude sequences overlap and do not indicate any significant age difference.

The presence of lithium has often been used to differentiate between the lithium-rich members of a young cluster and old lithium-depleted background stars. However, the Orion region is populated with many young groups of stars that may still show lithium but are not truly σ Ori cluster members and contaminate our sample (e.g. from the OBP-near group discussed in Section 4.4). We nevertheless prepared lithium data to reduce the possibility of background contamination. We thus performed the lithium test of youth to search for lithium-depleted stars that might be non-members, depending on their colour. Figure 9 shows Li pseudo-equivalent widths (pEWs) described by, for example, Pavlenko et al. (2007) for late-K and M dwarfs in both clusters. We note that measured pEW(Li) in veiled stars is subject to high variability.

According to Zapatero Osorio et al. (2002), the majority of the members in the σ Ori cluster are considered too young to exhibit depleted lithium. However, some exceptions have been observed, suggesting the presence of a few members with depleted lithium, which might indicate an older age (Kubiak et al. 2017). For instance, Sacco et al. (2008) reported approximately 25 members of σ Ori with a pEW(Li) of less than 150 mÅ.

In our sample, the distribution of lithium indicates a prevalence of very young Li-undepleted stars. However, there are three stars that are not veiled and have no (or almost no) lithium left in their atmospheres. They are all reliable members according to their *c* value. We plan to address the question of lithium depletion and possible age spread in the cluster with additional spectroscopic observations in our subsequent paper.

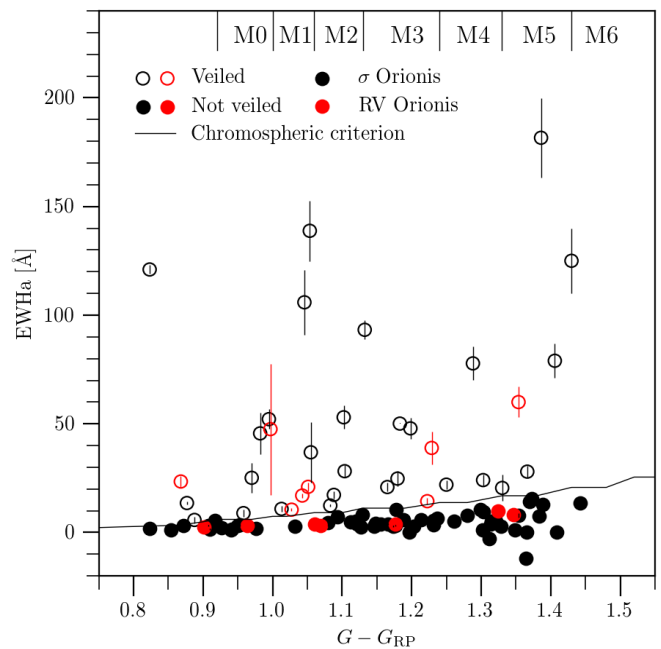

**Fig. 8.** Hα emission. Stars with very high emission (empty dots) are affected by veiling. The chromospheric criterion (solid line) to distinguish between the veiled and non-veiled stars follows Barrado y Navascués & Martín (2003) and is described in Appendix B.

Last but not least, the last spectroscopic indicator of youth considered in this work is the Na I subordinate doublet in the far red part of the optical spectrum. The equivalent width of Na I is known to be sensitive to surface gravity in M-type objects, and it has been used as an indicator of young age (Martín et al. 2010).

Stellar radius changes fast in young contracting stars in Hayashi tracks, and thus surface gravity can be used as a gravity indicator to distinguish between stars above and on the main sequence. Figure 10 reveals a colour-dependent trend of pEW(Na), with a few outliers displaying less sodium than the rest of the sample. Similar to lithium, sodium is also subject to variability due to veiling, but to a lesser extent because it is located at a longer wavelength. Most of the outliers are veiled, except stars a, b, c, and d, as annotated in Figure 10. Stars c and b are on the cluster sequence in the colour-magnitude diagram, raising questions as to why their sodium is low, and whether they are truly veiled despite their Hα being low. Active accretion may be episodic. Their Hα and Na measurements were not obtained simultaneously, and thus it may occur that they were quiet at the epoch when Hα was observed, but veiled when Na was observed.

Star a appears to be overluminous with respect to the cluster sequence, which is consistent with its small pEW(Na). These observations indicate an age that is younger than the two clusters. Although only setting an upper age limit, its undepleted lithium (0.51 ± 0.06 Å) speaks in favour of its very young age. This star is a reliable kinematic member ($c = 3.4$, although there is no radial velocity measurement), so it must have formed at the end of the Sigma-Orionis star-formation event. The existence of a few members much younger and much older than the mean age of the σ Ori cluster could provide important information about the star formation history.

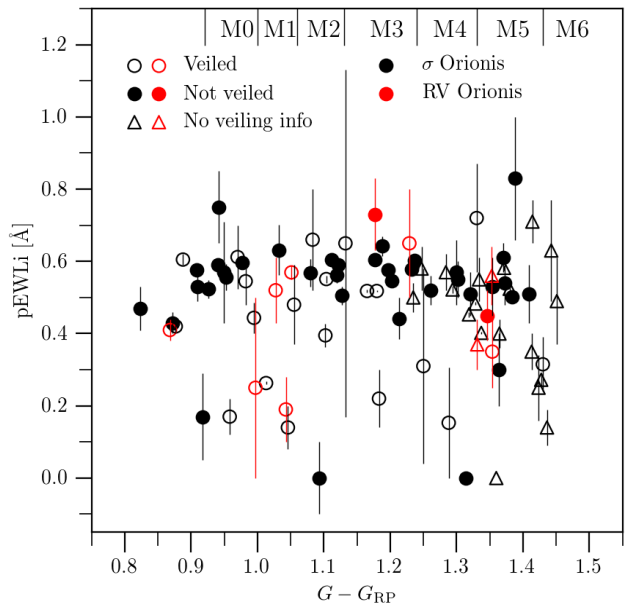

**Fig. 9.** Pseudo-equivalent width of lithium for $\sigma$ and RV Ori members. Most of the stars have the cosmic amount of lithium left. A large fraction of the members are affected by veiling, which causes variation in the measured pEW(Li). Among the stars that are not veiled, we found two stars with completely depleted lithium and one star with semi-depleted lithium.

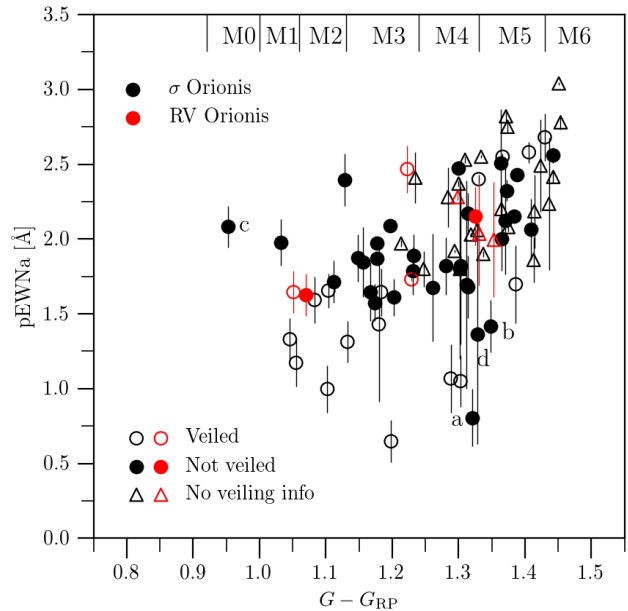

**Fig. 10.** Pseudo-equivalent width of sodium in the Sigma and RV Ori clusters. There is a clear dependence of pEW(Na) on colour. The presence of veiling can affect the measured pEW(Na). Outlier a is likely younger than the rest of the sample because it shows weaker Na equivalent widths and is more overluminous than the rest of the stars in $\sigma$ Ori with similar colours.

### 4.6. Comparison with the literature, contamination, and completeness

The contamination rate is about 12% in $\sigma$ Ori, based on the number of stars that are not found on the cluster sequence of the colour-magnitude diagram. The majority of these outliers are located in the outskirts of the cluster, so the contamination is likely smaller in the cluster's core.

We would like to emphasize that we refrained from making any cuts in parallaxes, proper motions, or magnitudes, as our primary goal was to achieve maximum completeness in the low-mass range. For almost 90% of the targets, the relative parallax errors are less than 10%, and they remain below 20% even for the faintest targets with G ~ 20. At this limit in *Gaia*, the completeness ratio for a 5 and 6-parameter solution is 92.2% (Lindegren et al. 2021).

Most of the existing membership analyses were conducted before *Gaia*'s era. Many of these studies primarily focused on the low-mass members or were part of large-scale investigations (e.g. Kounkel et al. 2018). However, our current work benefits from *Gaia*'s precision measurements, allowing us to delve into the fine structure of this cluster and conduct a more detailed study. In Table 2 we compare our new membership list with the literature. The comparison is based only on the stars from the literature that were also included in our input catalogue. We took all members from $\sigma$ Ori, RV Ori, and the Flame association into account. The comparison shows 50-85% agreement with the previous works. There are 82 stars in $\sigma$ Ori, 4 in RV Ori and 10 in the Flame association that are new members and are not found in any other catalogue mentioned here. They are mostly located in the outskirts of the cluster. On the other hand, there are nine stars that are listed as members in three or more reference catalogues and rejected in this work. They have high membership probabilities, but are located beyond 5 tidal radii.

### 5. Conclusions

Sigma Orionis is an important benchmark cluster in the field of stellar and substellar star formation and evolution due to its youth. However, it is located in a complex star-forming site and is thus challenging to isolate. In this work we used high-precision astrometry from *Gaia* DR3 to re-evaluate its membership using the modified convergent point algorithm. We explored the fine structure of this young star-forming region and described $\sigma$ Ori, RV Ori, and the Flame association. In total, we report 96 members that had never before been listed in the literature.

We supported their membership and young status with spectroscopic indicators, such as the equivalent widths of lithium, sodium, and H$\alpha$. Interestingly, RV Ori exhibits a higher proportion of veiled stars (53%) than $\sigma$ Ori (33%). On the other hand, their members lie on the same sequence of the colour-magnitude diagram. In future work, we plan to study the age distribution in this complex region by using several indicators (isochrones, lithium depletion, and surface gravity).

Knowledge about the substructure within complex star-forming regions plays a crucial role in stellar astrophysics, particularly when studying the initial mass function. However, in the substellar regime, the data are relatively sparse, leading to less well-constrained results. In this context, $\sigma$ Ori proves to be highly advantageous for such studies. Its youth, minimal extinction, and relative proximity to the Sun make it particularly convenient. Furthermore, the upcoming *Euclid* space mission will observe $\sigma$ Ori, providing valuable insights into substellar objects, extending down to planetary-mass objects. This mission

| Reference | Total | Overlap | Confirmed | Rejected |
|---|---|---|---|---|
| Sacco et al. (2008) | 62 | 47 | 40 (85%) | 7 (14%) |
| Zapatero Osorio et al. (2002) | 27 | 20 | 16 (80%) | 4 (20%) |
| Caballero et al. (2019) | 103 | 60 | 46 (76%) | 14 (23%) |
| Hernández et al. (2014) | 238 | 128 | 93 (72%) | 35 (27%) |
| Maxted et al. (2008) | 237 | 78 | 49 (62%) | 29 (37%) |
| Kenyon et al. (2005) | 70 | 40 | 21 (52%) | 19 (47%) |

**Notes.** We list the total number of stars in the catalogues and the overlap with our input catalogue. Among the objects in the overlap, we counted the number of stars we confirmed and rejected as members in this work. Some of these rejected members have high membership probability but are located beyond 5 tidal radii from the cluster centre.

**Table 2.** Comparison with the literature.

promises to enhance our understanding of these objects in the cluster.

*Acknowledgements.* We would like to express our gratitude to the anonymous referee, whose detailed reading and constructive feedback enhanced the clarity and quality of our work.
Funding for MZ and ELM was provided by the European Union (ERC Advanced Grant, SUBSTELLAR, project number 101054354). APG acknowledges support from the grant PID2020-120052GB-I00 financed by MCIN/AEI/10.13039/501100011033. This work has made use of data from the European Space Agency (ESA) mission *Gaia* (https://www.cosmos.esa.int/gaia), processed by the *Gaia* Data Processing and Analysis Consortium (DPAC, https://www.cosmos.esa.int/web/gaia/dpac/consortium). Funding for the DPAC has been provided by national institutions, in particular the institutions participating in the *Gaia* Multilateral Agreement. This publication makes use of VOSA, developed under the Spanish Virtual Observatory (https://svo.cab.inta-csic.es) project funded by MCIN/AEI/10.13039/501100011033/ through grant PID2020-112949GB-I00. VOSA has been partially updated by using funding from the European Union's Horizon 2020 Research and Innovation Programme, under Grant Agreement nº 776403 (EXOPLANETS-A). This research has made use of the Simbad and Vizier databases, and the Aladin sky atlas operated at the centre de Données Astronomiques de Strasbourg (CDS), and of NASA's Astrophysics Data System Bibliographic Services (ADS). The Digitized Sky Survey was produced at the Space Telescope Science Institute under U.S. Government grant NAG W–2166. The images of these surveys are based on photographic data obtained using the Oschin Schmidt Telescope on Palomar Mountain and the UK Schmidt Telescope. The plates were processed into the present compressed digital form with the permission of these institutions. Software: `astropy` (Price-Whelan et al. 2018), `NumPy` (Harris et al. 2020), `IPython` (Pérez & Granger 2007), `TOPCAT` (Taylor 2005) and `matplotlib` (Hunter 2007).

# Appendix A: Membership list

Table A.1. Membership list.

| source_id | ra | dec | $\mu_\alpha^*$ | $\mu_\delta$ | $\pi$ | $\sigma_\pi$ | RV | $\sigma_{RV}$ | refRV | ruwe | g | g_rp | memb | c | outlier |
|---|---|---|---|---|---|---|---|---|---|---|---|---|---|---|---|
| Gaia DR3 | deg | deg | mas yr$^{-1}$ | mas yr$^{-1}$ | mas | mas | km s$^{-1}$ | km s$^{-1}$ | | | mag | mag | | | |
| 3216486443742786048 | 84.686522 | −2.600079 | 22.63[a] | 13.45[a] | 2.58[b] | 0.01[b] | | | | | 3.8 | −0.2 | σ Ori | 11.36 | |
| 3216110234671579648 | 84.609325 | −2.678168 | 1.28 | −0.54 | 2.57 | 0.23 | 32.1 | 0.4 | 1 | 1.03 | 18.6 | 1.4 | σ Ori | 1.43 | |
| 3217572555072526464 | 83.549769 | −1.348554 | 1.19 | −0.35 | 2.43 | 0.02 | 28.0 | 5.0 | 1 | 1.16 | 13.0 | 0.7 | σ Ori | 11.79 | |
| 3217400206624439808 | 85.060094 | −1.733046 | 1.67 | −0.63 | 2.43 | 0.02 | 32.7 | 2.0 | 1 | 1.08 | 11.2 | 0.5 | σ Ori | 1.62 | CMD |
| 3216109861009561728 | 84.684317 | −2.672163 | 1.46 | −1.01 | 2.47 | 0.01 | 32.8 | 7.0 | 1 | 1.03 | 13.3 | 0.9 | σ Ori | 3.39 | |
| 3216494793158602368 | 84.947584 | −2.437867 | 1.93 | −0.70 | 2.39 | 0.05 | 34.9 | 5.1 | 1 | 2.95 | 11.6 | 0.7 | σ Ori | 2.28 | |
| 3216486203224073088 | 84.714404 | −2.605735 | 0.99 | −0.19 | 2.38 | 0.13 | 32.1 | 0.7 | 2 | 2.73 | 15.9 | 1.4 | σ Ori | 4.64 | |
| 3216557873341545216 | 85.845124 | −2.269162 | 1.26 | −0.14 | 2.52 | 0.04 | 56.2 | 7.3 | 1 | 1.53 | 14.6 | 1.1 | σ Ori | 11.49 | RV |
| 3216549631300819072 | 85.192372 | −1.926771 | 1.95 | −1.55 | 2.59 | 0.04 | 18.1 | 3.4 | 1 | 1.68 | 9.6 | 0.0 | σ Ori | 11.92 | CMD |
| 3216488335527834880 | 84.771867 | −2.550167 | 1.72 | −0.40 | 2.53 | 0.06 | 36.0 | 3.0 | 1 | 1.07 | 16.4 | 1.2 | σ Ori | 0.98 | |
| 3216108933296640640 | 84.652893 | −2.737151 | 1.49 | −1.06 | 2.48 | 0.05 | 37.1 | 7.0 | 1 | 1.05 | 15.8 | 1.3 | σ Ori | 6.28 | |
| 3216486271943551616 | 84.702870 | −2.604529 | 1.34 | −0.44 | 2.45 | 0.03 | 32.5 | 0.5 | 2 | 1.13 | 14.9 | 1.1 | σ Ori | 1.29 | |
| 3216439057367561472 | 84.893050 | −2.646377 | 1.53 | −1.25 | 2.51 | 0.25 | 31.0 | 0.4 | 1 | 1.05 | 18.6 | 1.4 | σ Ori | 3.66 | |
| 3216446380287243392 | 85.021907 | −2.514555 | 2.18 | −0.54 | 2.54 | 0.11 | 31.5 | 0.4 | 1 | 0.93 | 17.4 | 1.3 | σ Ori | 3.83 | |
| 3216486924778686976 | 84.696624 | −2.576921 | 1.09 | 0.02 | 2.55 | 0.06 | 31.9 | 1.1 | 2 | 1.24 | 16.1 | 1.3 | σ Ori | 9.15 | |
| 3216497988614248832 | 84.983115 | −2.343545 | 1.78 | −1.41 | 2.51 | 0.08 | 20.1 | 10.2 | 1 | 3.17 | 15.0 | 1.1 | σ Ori | 9.52 | |
| 3023970650732518272 | 86.167877 | −3.587717 | 1.41 | −0.49 | 2.53 | 0.04 | 24.6 | 5.2 | 1 | 0.98 | 15.8 | 1.3 | σ Ori | 0.37 | |
| 3216486233291778944 | 84.701223 | −2.611411 | 1.44 | −0.17 | 2.51 | 0.04 | 30.2 | 1.4 | 2 | 1.22 | 15.4 | 1.3 | σ Ori | 3.44 | |
| 3217319839196766592 | 84.484609 | −1.715423 | 1.04 | −0.53 | 2.40 | 0.03 | 54.2 | 6.0 | 1 | 1.00 | 14.7 | 1.0 | σ Ori | 10.68 | RV |
| 3217293004241142528 | 84.403117 | −1.848256 | 1.19 | −0.69 | 2.39 | 0.03 | −8.7 | 8.1 | 1 | 1.04 | 15.0 | 1.0 | σ Ori | 11.60 | RV |
| 3216431468160892032 | 85.213457 | −2.599590 | 2.67 | −0.28 | 2.44 | 0.08 | | | | 1.11 | 17.0 | 1.2 | RV Ori | 3.14 | CMD |
| 3216501527667171712 | 84.861712 | −2.323555 | 2.03 | −0.13 | 2.50 | 0.02 | 38.5 | 7.8 | 1 | 1.07 | 14.6 | 1.1 | RV Ori | 2.74 | |
| 3216467408447094400 | 85.242317 | −2.525601 | 2.82 | −0.41 | 2.55 | 0.03 | 3.4 | 5.2 | 1 | 1.15 | 15.1 | 1.1 | RV Ori | 11.91 | RV |
| 3216497713736341248 | 85.008160 | −2.359065 | 2.34 | 0.02 | 2.51 | 0.04 | 31.2 | 0.3 | 1 | 1.01 | 15.7 | 1.2 | RV Ori | 1.25 | |
| 3216524720490689024 | 84.997798 | −2.281205 | 2.40 | −0.08 | 2.54 | 0.05 | 31.6 | 0.3 | 1 | 1.14 | 16.1 | 1.3 | RV Ori | 0.68 | |
| 3216501871264551040 | 84.870155 | −2.297600 | 2.03 | −0.11 | 2.49 | 0.04 | | | | 1.01 | 15.8 | 1.2 | RV Ori | 1.78 | |
| 3216496408065080448 | 84.976145 | −2.411151 | 2.25 | −0.10 | 2.45 | 0.03 | | | | 1.21 | 14.7 | 1.2 | RV Ori | 0.08 | |
| 3216498022973982336 | 85.004239 | −2.333289 | 2.29 | 0.08 | 2.54 | 0.06 | 30.8 | 0.9 | 1 | 1.06 | 16.6 | 1.4 | RV Ori | 1.51 | |
| 3216494793158603392 | 84.944216 | −2.442052 | 1.50 | 0.17 | 2.47 | 0.06 | | | | 1.06 | 16.4 | 1.3 | RV Ori | 10.42 | |
| 3216538086428775040 | 84.887210 | −2.191991 | 1.87 | 0.40 | 2.49 | 0.04 | | | | 1.00 | 16.1 | 1.2 | RV Ori | 6.87 | |
| 3216496476785764736 | 85.038907 | −2.418559 | 2.23 | −0.39 | 2.50 | 0.07 | 31.6 | 0.3 | 1 | 1.08 | 16.5 | 1.3 | RV Ori | 2.09 | |

**Notes.** Proper motions, parallaxes, ruwe, g (phot_g_mean_mag) and g_rp are propagated from *Gaia* DR3; references for radial velocities are listed in refRV. The entire table for all clusters in this work is available only in electronic form at the CDS via anonymous ftp to cdsarc.cds.unistra.fr (130.79.128.5) or via https://cdsarc.cds.unistra.fr/cgi-bin/qcat?J/A+A/.
[a] Proper motion from Hipparcos (van Leeuwen 2007).
[b] Interferometric parallax (Schaefer et al. 2016).

**References.** (1) *Gaia* DR3, (2) Sacco et al. (2008).

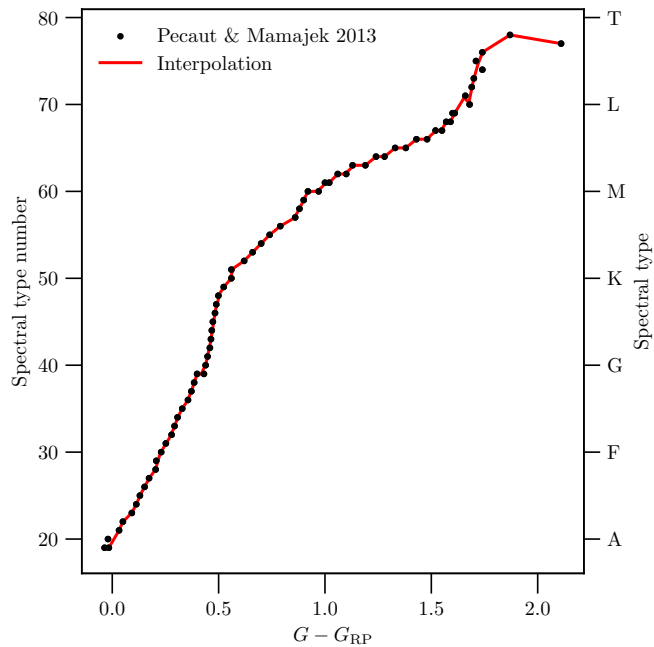

**Fig. B.1.** Relation between the colour and spectral type from Pecaut & Mamajek (2013) as described in Appendix B. We used a linear interpolation to assign a spectral type number (following the convention from Barrado y Navascués & Martín 2003) from the $G - G_{RP}$ colour of the star.

## Appendix B: Veiling

Very young stars – for example, T Tauri objects – are subject to veiling that results from the interaction with their active accretion discs. These processes can alter the continuum level of the stellar spectrum as well as their spectral lines (e.g. Rei et al. 2018). Barrado y Navascués & Martín (2003) provided an empirical chromospheric criterion to distinguish between the veiled and non-veiled stars, depending on their spectral type. Their relation is described with

$$\log_{10} \mathrm{EW}(\mathrm{H}\alpha) = 0.0893\,(\mathrm{Sp.\ Type}) - 4.5767, \tag{B.1}$$

where the spectral type is given in a numeric form; for example, O1 corresponds to 1, B1 to 11, A1 to 21, and so on. This relation is valid between spectral types G5 and M5.5.

Since our data table lacks spectral types for some stars, we used *Gaia*'s $G - G_{RP}$ colours instead. Conveniently, we used an updated table[3] originally published in Table 5 by Pecaut & Mamajek (2013) and interpolated the data to estimate the relationship between the colour and spectral types, as illustrated in Figure B.1. The chromospheric criterion in the $G - G_{RP}$ versus EW(H$\alpha$) diagram is shown in Figure 8. In Table C.1 we gathered the information on H$\alpha$ and estimate the presence of veiling.

## Appendix C: Spectroscopic youth indicators

We prepared a compilation of spectroscopic data to further confirm the youth of the cluster members. We obtained pEW(Li) 6708 Å measurements from Caballero et al. (2019), Sacco et al. (2008), Kenyon et al. (2005), Zapatero Osorio et al. (2002), and Hernández et al. (2014), pEW(Na) measurements from Kenyon et al. (2005), Hernández et al. (2014), and Maxted et al. (2008), and H$\alpha$ equivalent width measurements from Caballero et al. (2019), Kenyon et al. (2005), and Hernández et al. (2014). An extract of the compilation is presented in Table C.1; the entire table for all stars with available spectroscopy is available at the CDS.

For a few stars we found spectroscopic measurements in more than one paper (maximum three). In such cases, we listed the average value and determined the uncertainty as the standard deviation. Uncertainties for all equivalent widths in Hernández et al. (2014) were estimated to be 10%. We set the errors accordingly, and used 0.1 Å when the measured equivalent width equals 0.

Hernández et al. (2014) measured the equivalent width of the sodium line at 8195 Å while the rest of the sources cover both lines of the doublet (8183 and 8195 Å). The ratio between the two lines in this Na I doublet is constant, and we determined the correction factor of $1.75 \pm 0.16$ that is used to estimate the equivalent width of the whole doublet. This correction was obtained by us using the routine `splot` in the Image Reduction and Analysis Facility (IRAF) environment (Tody 1986). We compared the equivalent width of the Na I doublet using the direct integration option of the blended doublet, and the individual equivalent width of each line in the doublet using the Gaussian de-blending option. The correction was found to be insensitive to spectral type and spectral resolution using the collection of spectra of M-type stars presented in Martín et al. (2010).

---

[3] `http://www.pas.rochester.edu/~emamajek/EEM_dwarf_UBVIJHK_colors_Teff.txt`, v2021.03.02.

**Table C.1.** Spectroscopic indicators and spectral types (extract).

| source_id Gaia DR3 | pEW(Li) Å | epEW(Li) Å | refLi | EW(Hα) Å | eEW(Hα) Å | refHα | veiled | pEW(Na) Å | epEW(Na) Å | refNa | SpT | refSpT |
|---|---|---|---|---|---|---|---|---|---|---|---|---|
| 3216486478101979520 | 0.57 | 0.02 | 2,3 | 21.0 | 2.0 | 3,7 | Y | 1.6 | 0.1 | 7 | M1.0 | 2 |
| 3216500531234897920 | 0.41 | 0.03 | 3 | 23.4 | 2.4 | 3,7 | Y | | | | K4 | 3 |
| 3216447239280699008 | | | | 3.1 | 0.3 | 7 | N | | | | K7.0 | 7 |
| 3216524651771210240 | 0.52 | 0.09 | 3 | 10.5 | 0.7 | 3 | Y | | | | M0 | 3 |
| 3216521765553185152 | 0.19 | 0.09 | 3 | 17.0 | 1.0 | 3 | Y | | | | M1.5 | 3 |
| 3216446689524900480 | | | | 2.6 | 0.3 | 7 | N | | | | K7.0 | 7 |
| 3216500634314117504 | | | | 3.7 | 0.4 | 7 | N | | | | M1.0 | 7 |
| 3216446002330124416 | 0.45 | 0.07 | 3 | 8.1 | 0.6 | 3 | N | | | | M7 | 3 |
| 3216496030109176448 | 0.73 | 0.10 | 6 | 4.0 | 0.8 | 6 | N | | | | M3 | 6 |
| 3216445899250910976 | 0.25 | 0.25 | 3,7 | 47.5 | 30.1 | 3,7 | Y | | | | K7 | 3 |
| 3216497267059750912 | 0.35 | 0.10 | 6 | 60.0 | 7.0 | 6 | Y | | | | M5.5 | 6 |
| 3216501527667171712 | | | | 3.3 | 0.3 | 7 | N | 1.6 | 0.1 | 7 | M1.5 | 7 |
| 3216497713763412248 | 0.65 | 0.15 | 6 | 38.9 | 7.6 | 6,7 | Y | 1.7 | 0.0 | 5 | M4 | 6 |
| 3216524720490689024 | | | | | | | | 2.3 | 0.0 | 5 | | |
| 3216501871264551040 | | | | 14.4 | 1.4 | 7 | Y | 2.5 | 0.2 | 7 | M3.5 | 7 |
| 3216498022973982336 | 0.56 | 0.08 | 4 | | | | | 2.0 | 0.4 | 4,5 | | |
| 3216494793158603392 | | | | 9.9 | 1.0 | 7 | N | 2.2 | 0.2 | 7 | M5.0 | 7 |
| 3216496476785764736 | 0.37 | 0.07 | 4 | | | | | 2.0 | 0.3 | 4,5 | | |
| 3216110234671579648 | 0.51 | 0.08 | 4 | | | | N | 2.1 | 0.2 | 4,5 | M4.5 | 7 |
| 3216109861009561728 | 0.61 | 0.02 | 3,6 | 5.8 | 1.6 | 3,6,7 | Y | | | | K8 | 6 |
| 3216494793158602368 | 0.44 | 0.05 | 3 | 1.5 | 0.2 | 3 | N | | | | K0 | 3 |
| 3216486203224073088 | | | | | | | | | | | K9.5 | 2 |
| 3216488335327834880 | 0.58 | 0.06 | 4 | | | | | 1.8 | 0.1 | 4 | M4.5 | 2 |
| 3216108933296640640 | 0.57 | 0.00 | 2 | | | | | 1.8 | 0.0 | 5 | K9.5 | 2 |
| 3216486271943551616 | 0.59 | 0.00 | 2 | 3.8 | 0.4 | 7 | N | | | | | |
| 3216439057367561472 | 0.63 | 0.14 | 4 | | | | | 2.4 | 0.0 | 4,5 | M5.5 | 6 |
| 3216446380287243392 | 0.72 | 0.15 | 6 | 20.5 | 6.0 | 6 | Y | 2.4 | 0.0 | 5 | M3.5 | 2 |
| 3216486924778686976 | 0.15 | 0.15 | 2,7 | 77.8 | 7.8 | 7 | Y | 1.1 | 0.2 | 7 | M2.0 | 7 |
| 3216497988614248832 | | | | 8.2 | 0.8 | 7 | N | 2.4 | 0.2 | 7 | M4.5 | 2 |
| 3216486233291778944 | 0.51 | 0.06 | 2 | 5.2 | 0.5 | 7 | N | 0.8 | 0.2 | 7 | | |
| 3216109723570616320 | 0.55 | 0.06 | 4 | | | | | 2.5 | 0.0 | 4,5 | | |
| 3216442355903486080 | 0.31 | 0.07 | 3,4 | 125.0 | 15.0 | 3 | Y | 2.7 | 0.2 | 4,5 | M6.5 | 3 |
| 3216111952658344064 | 0.61 | 0.04 | 4 | 14.3 | 1.4 | 7 | N | 2.1 | 0.4 | 4,5,7 | M5.5 | 7 |
| 3216112919026100352 | 0.54 | 0.06 | 4 | 15.5 | 1.6 | 7 | N | 2.3 | 0.1 | 4,5,7 | M5.0 | 7 |

**Notes.** If measurements are available in multiple catalogues, we computed the mean value and standard deviation. Corresponding references are provided for each catalogue. Chromospheric criterion from Barrado y Navascués & Martín (2003) is used to distinguish between veiled (Y) and non-veiled (N) stars. The entire table for all stars with available spectroscopy is available only in electronic form at the CDS via anonymous ftp to cdsarc.cds.unistra.fr (130.79.128.5) or via https://cdsarc.cds.unistra.fr/cgi-bin/qcat?J/A+A/.

**References.** (2) Sacco et al. (2008), (3) Caballero et al. (2019), (4) Kenyon et al. (2005), (5) Maxted et al. (2008), (6) Zapatero Osorio et al. (2002), (7) Hernández et al. (2014).